\documentclass{phbauth}
\usepackage{graphicx}

\begin{document}
\begin{frontmatter}

\title{Magnetoresistance and conductivity exponents of quench-condensed 
ultra-thin films of Bi}

\author{K. Das Gupta\thanksref{thank1}},
\author{G. Sambandamurthy},
\author{N. Chandrasekhar}

\address{Department of Physics, Indian Institute of 
 Science, Bangalore 560 012, India}

\thanks[thank1]{ E-mail: kantimoy@physics.iisc.ernet.in }

\begin{abstract}
We have studied the magnetoresistance (MR) and evolution of conductivity
with  thickness  of  quench-condensed  Bismuth  films  on  substrates of
various  dielectric constants. Our results indicate a negative intial MR
proportional  to the square of the magnetic field. The conductance shows
a power-law  kind of  dependence on thickness, with an exponent close to
1.33,  characterisitic  of a 2-D percolating system, only when the films
are  grown  on a thin  ($\sim 10${\rm \AA} Germanium underlayer but not
otherwise.

\end{abstract}

\end{frontmatter}

\section{Introduction}
Ultra-thin films of metals, typically less than $\sim 50${\rm\AA} thick, 
quench-condensed on to substrates held at $20$K or lower have served as 
models of 2-D disordered systems. Although the first {\it in-situ}
studies were done in the fifties\cite{BUCKEL}, recent use of the techniques of MBE  
to grow these films in a clean UHV environment have yielded new  
results\cite{GOLDMAN1,GOLDMAN2}.

\section{Experimental }
We have studied the MR and evolution of the sheet
conductivity($\sigma$) of ultra-thin quench-condensed Bi films on 
a-quartz and sapphire, with and without pre-deposited underlayers 
of Ge and Sb. The dielectric constant of 
the substrates range from about $3$ for Sapphire to nearly $15$ for 
Ge. The dielectric constant increases when a polarisable  underlayer is 
present, however the exact change, due to a $\sim 10${\rm \AA}underlayer is difficult
to predict. The metal was evaporated from a Knudsen-cell source  
at a steady rate of $\sim 5${\rm\AA}/min. 
The substrate was cooled to $\sim 20$K. The cryostat 
equipped with a superconducting magnet was pumped by a turbomolecular pump,
backed by a diaphragm pump. Hydrocarbon free vacuum of
$\sim 5\times10^{-8}$ Torr, was maintained throughout.

\section{Results}
 
\begin{figure}[h]
\begin{center}\leavevmode
\includegraphics[width=1.0\linewidth,clip=]{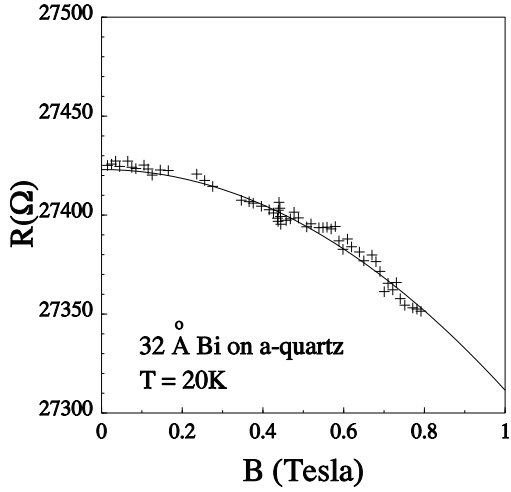}
\caption{ Decrease of the sheet resistance of a film with 
magnetic field. The line is a best fit, of the form 
$R(B)=R(0) - k B^2$, where $k$ is some constant.
}\label{mrplot}\end{center}\end{figure}

In Fig \ref{mrplot} we have shown the transverse MR of a Bi film quench condensed
on quartz. For normal metals the semiclassical model of electron motion 
predicts a positive MR, though the exact form is complicated and depends 
among other things on the orientation of the direction of the field and 
current with  respect to the open electron orbits\cite{ASHCROFT}.
 A negative MR, at low
temperatures can however result from weak localisation effects \cite{BERGMAN}. A negative 
MR proportional to the square of the field has not been reported 
earlier in these systems. \\

\begin{figure}[btp]
\begin{center}\leavevmode
\includegraphics[width=1.0\linewidth, height=1.0\linewidth]{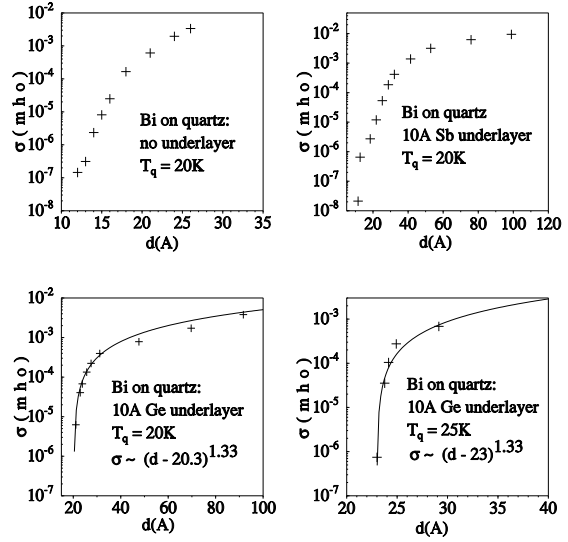}
\caption{Evolution of conductivity with thickness of the film. $T_q$ denotes
the substrate temperature during deposition. 
}\label{resvsdplot}\end{center}\end{figure}

In most studies of these systems an underlayer of Sb/Ge has been used by 
experimenters. Fig \ref{resvsdplot} shows the evolution of $\sigma$ as the nominal film 
thickness $d$) is increased in four  Bi films grown under different conditions. 
Only the films on Ge underlayer seem to be following a power law of the form 
$\sigma \sim (d-d_0)^\nu$, with $\nu$ close to $1.33$. This behaviour is 
well-known for a 2-D random resistor network close to its percolation threshold
\cite{STAUFFER}.

 Localisation is a quantum mechanical effect whereas percolation is generally
seen as a classical phenomenon. The interplay between percolation and 
localisation makes the question of conduction mechanism in these films
a very interesting one.\\

\begin{ack}
This work is supported by Department of Science and Technology, Govt. of 
India. KDG would like to thank Council for Scientific
and Industrial Research, India, for a junior research fellowship.
\end{ack}


\begin{thebibliography}{9}
\bibitem{BUCKEL}Buckel, W. and Hilsch, R. (1954), Z. Phys. {\bf 138}, 109.
\bibitem{GOLDMAN1} Orr, B. G. (1985), Doctoral Dissertation, University of Minnesota.
\bibitem{GOLDMAN2}Haviland, D.B. Liu, Y. and Goldman, A.M. (1989), Phys. Rev. Lett.
 {\bf 62}, 2180. 
\bibitem{ASHCROFT}N. W. Ashcroft \& D. Mermin, Solid State Physics, pp234,
Saunder's College Publishing, 1976.
\bibitem{BERGMAN}G. Bergman, (1984), Physics Reports, {\bf 107}, 1.
\bibitem{STAUFFER}D. Stauffer, (1979),  Physics Reports, {\bf 54}, 1.  
\end{thebibliography}
\end{document}